\documentclass[pra,showkeys,showpacs,groupedaddress,twocolumn]{revtex4}
\usepackage{amssymb}
\usepackage{graphicx}
\usepackage{dcolumn}
\usepackage{bm}
\usepackage{amsmath}
\usepackage{subfigure}
\usepackage{float}
\usepackage{color}

\begin{document}
\title{Spectroscopy of cesium Rydberg atoms in strong radio-frequency fields}

\author{Yuechun Jiao$^{1,3}$}
\author{Zhiwei Yang$^{1,3}$}
\author{Jingkui Li$^{1,3}$}
\author{Georg Raithel$^{1,2}$}
\author{Jianming Zhao$^{1,3}$}
\thanks{Corresponding author: zhaojm@sxu.edu.cn}
\author{Suotang Jia$^{1,3}$}

\affiliation{$^{1}$State Key Laboratory of Quantum Optics and Quantum Optics Devices, Institute of Laser Spectroscopy, Shanxi University, Taiyuan 030006, China}
\affiliation{$^{2}$ Department of Physics, University of Michigan, Ann Arbor, Michigan 48109-1120, USA}
\affiliation{$^{3}$Collaborative Innovation Center of Extreme Optics, Shanxi University, Taiyuan 030006, China}
\date{\today}

\begin{abstract}
We study Rydberg atoms modulated by strong radio-frequency (RF) fields with a frequency of 70~MHz.
The Rydberg atoms are prepared in a room temperature cesium cell, and their level structure is probed using electromagnetically induced transparency (EIT). As the RF field increases from the weak- into the strong-field regime, the range of observed RF-induced phenomena progresses from AC level shifts through increasingly pronounced and numerous RF-modulation sidebands to complex state-mixing and level-crossings with high-\emph{l} hydrogen-like states. Weak anharmonic admixtures in the RF field
generate clearly visible modifications in the Rydberg-EIT spectra. A Floquet analysis is employed to model the Rydberg spectra, and good agreement with the experimental observations is found.
Our results show that all-optical spectroscopy of Rydberg atoms in vapor cells can serve as an antenna-free, atom-based and calibration-free technique to measure and map RF electric fields and to analyze their higher-harmonic contents.
\end{abstract}
\keywords{Rydberg EIT, RF field modulation, Floquet theory}
\pacs{32.80.Rm, 42.50.Gy, 32.30.Bv}
\maketitle

\section{Introduction}
Atom-based field measurement has made significant progress in reproducibility, accuracy and resolution. Atoms have been successfully used for magnetometry with high sensitivity and spatial resolution~\cite{Savukov, Patton}. Rydberg atoms (highly excited atoms with principal quantum numbers $n\gg1$) have applications in electrometry due to their large DC polarizabilities and microwave-transition dipole moments, which follow respective scaling laws $\propto n^{7}$ and $\propto n^4$~\cite{Gallagher} and make these atoms extremely sensitive to DC and AC electric fields. Rydberg electromagnetically induced transparency (EIT)~\cite{Mohapatra} in atomic vapor cells has been used to realize a giant DC Kerr coefficient~\cite{K.Mohapatra} and to measure the electric fields of electromagnetic radiation with a large dynamic range~\cite{C. Holloway}. Radio frequency (RF)-dressed Rydberg EIT has been demonstrated in a number of applications, including measurements of microwave fields and polarizations~\cite{Sedlacek2012, A. Sedlacek}, millimeter waves~\cite{Gordon}, static electric fields~\cite{Barredo}, and precise determinations of quantum defects~\cite{Grimmel}.
RF measurement via Rydberg-EIT does not require  vacuum, atomic-beam and laser-cooling infrastructure, it offers significant potential for miniaturization~\cite{Budker}, and it covers a frequency range extending from the MHz into the THz-range.

To take advantage of Rydberg-EIT spectroscopy as an atom-based, calibration-free field measurement method, it is necessary to
theoretically model the experimentally produced spectra. In weak-field regimes, field measurement in the radio FM band (tens to hundreds of MHz) using Rydberg-EIT has been modeled well by perturbation theory~\cite{Bason}. In strong RF fields, the Rydberg levels exhibit higher-order couplings
and state-mixing with high-\emph{l} hydrogen-like states. In this case, the atom-field interaction can no
longer be modeled using perturbation theory. Typically, the RF field to be probed is periodic over the time scale of interest. Then, Floquet theory~\cite{Floquet, Floquet_Hydrogen}, a non-perturbative method akin to band structure theory, can be used to accurately describe the system.

In this work, we present EIT spectroscopy of Rydberg atoms modulated by strong RF fields with a frequency of 70~MHz. The dependence of the Rydberg spectrum on the strength of the RF field is investigated for harmonic and slightly distorted RF signals. The Rydberg spectra exhibit RF-field-induced level shifts and, in the strong RF-field regime, state-mixing with high-\emph{l} hydrogen-like states. We use the Floquet treatment to model the Rydberg-level shifts, state mixing and excitation rates. The RF field is deduced by comparison of experimental and calculated Floquet maps. We also find that the Rydberg spectrum exhibits different sideband structures for slightly distorted RF waveforms, demonstrating that the atomic response is sensitive to small anharmonic content of the applied RF signal.

\section{Experimental Setup}

\begin{figure}[ht]
\centering
\includegraphics[width=0.5\textwidth]{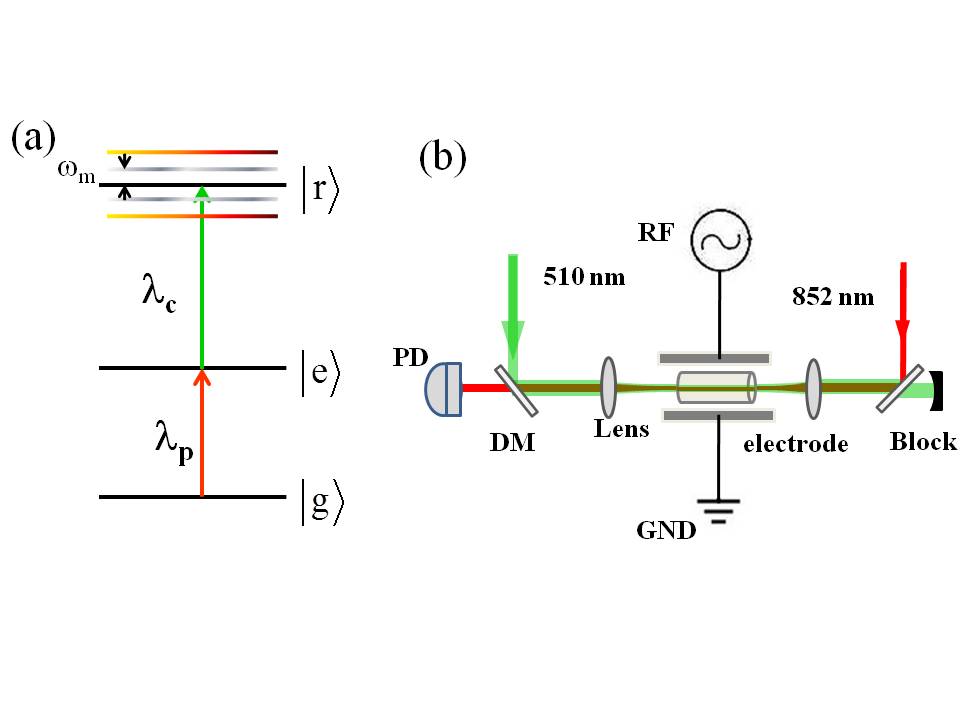}
\caption{(a) Energy level scheme of Rydberg-EIT. The probe laser, $\lambda_{p}$, is resonant with the transition $\vert g \rangle $
 $\to$ $\vert e \rangle $, and the coupling laser, $\lambda_{c}$, is scanned through resonances from $\vert e \rangle $ into Rydberg states $\vert r \rangle $. An applied RF electric field (frequency $\omega_{m}$ = 2$\pi \times$70~MHz) produces
field-mixed Rydberg levels with RF modulation sidebands, which are separated in energy by even
multiples of $\hbar \omega_{m}$. (b) Schematic of the experimental setup. The coupling and probe lasers counter-propagate through a cesium vapor cell.
The transmission of the probe beam through the cell is detected with a photodiode (PD).}
\end{figure}

A schematic of the experimental setup is shown in Fig.~1~(b). The experiments are performed in a room-temperature cesium cell. The relevant atomic levels form a three-level system, illustrated in Fig.~1~(a), consisting of the ground state 6S$_{1/2}$ (F = 4) ($|g\rangle$), the intermediate state 6P$_{3/2}$ (F'= 5) ($|e\rangle$) and the 57S$_{1/2}$ Rydberg state. A weak probe laser (wavelength $\lambda_{p}$ = 852~nm,  Rabi frequency $\Omega_p$ = 2$\pi \times $7.4~MHz, power = 1.5$\mu$W and $1/e^{2}$ waist $w_0 = 75~\mu$m) is resonant with the transition $|g\rangle \to |e\rangle$, while a strong coupling laser ($\lambda_{c}$ =510~nm, Rabi frequency $\Omega_c$ = 2$\pi \times $7.2~MHz, power=45~mW and $1/e^{2}$ waist
$w_0= 95~\mu$m) scans through the $|e \rangle \to |r\rangle$ Rydberg transitions, where $|r\rangle$ is the 57S$_{1/2}$ Rydberg level or an RF-dressed or RF-coupled Rydberg level close to it. The counter-propagating coupler and probe beams are linearly polarized, with polarizations parallel to each other.
The coupling laser results in an increased transmission of the probe laser when
EIT double-resonance condition is met. The EIT signal is observed by measuring the transmitted power of the probe beam.

The Rydberg level is modulated with a RF electric field with
modulation frequency $\omega_{m}$ = 2$\pi \times $70~MHz and variable field strength.
The RF field, provided by a function generator (Agilent 33250A), is applied to parallel-plate electrodes (spacing 27.0~mm), see Fig.~1~(b). The electric field between the two electrodes is uniform along the entire length of the atom-field interaction volume and parallel to the polarizations of the laser beams.
At low RF fields, the modulation generates an overall AC shift, because S-Rydberg levels of cesium exhibit quadratic Stark effect, as well as a sequence of modulation sidebands~\cite{Bason}, as sketched in Fig.~1~(a). In the strong-field domain, complex Rydberg-EIT spectra arise from strongly-mixed Floquet states and their RF sidebands.

We employ an auxiliary EIT setup without RF-field (not shown in Fig.~1) that provides a reference EIT spectrum, shown by the black line at the bottom of Fig.~2.
There, the probe laser is locked to the $|g\rangle \to |e\rangle$ transition, while the coupling laser is scanned
across the $|e\rangle \to |57 S_{1/2}\rangle$ transition. The main peak in the reference spectrum defines the position of 0 detuning.
The small peak red-detuned by 168~MHz from the main peak, marked with a red circle, is the EIT signal due to the hyperfine component $F'= 4$ of the intermediate 6P$_{3/2}$-level. (Note a Doppler factor~$\lambda_p/\lambda_c - 1 = 0.67$ applies~\cite{Rapol}.)

\section{Rydberg-EIT measurements in strong RF fields}

 \begin{figure}[thbp]
\centering
\includegraphics[width=0.5\textwidth]{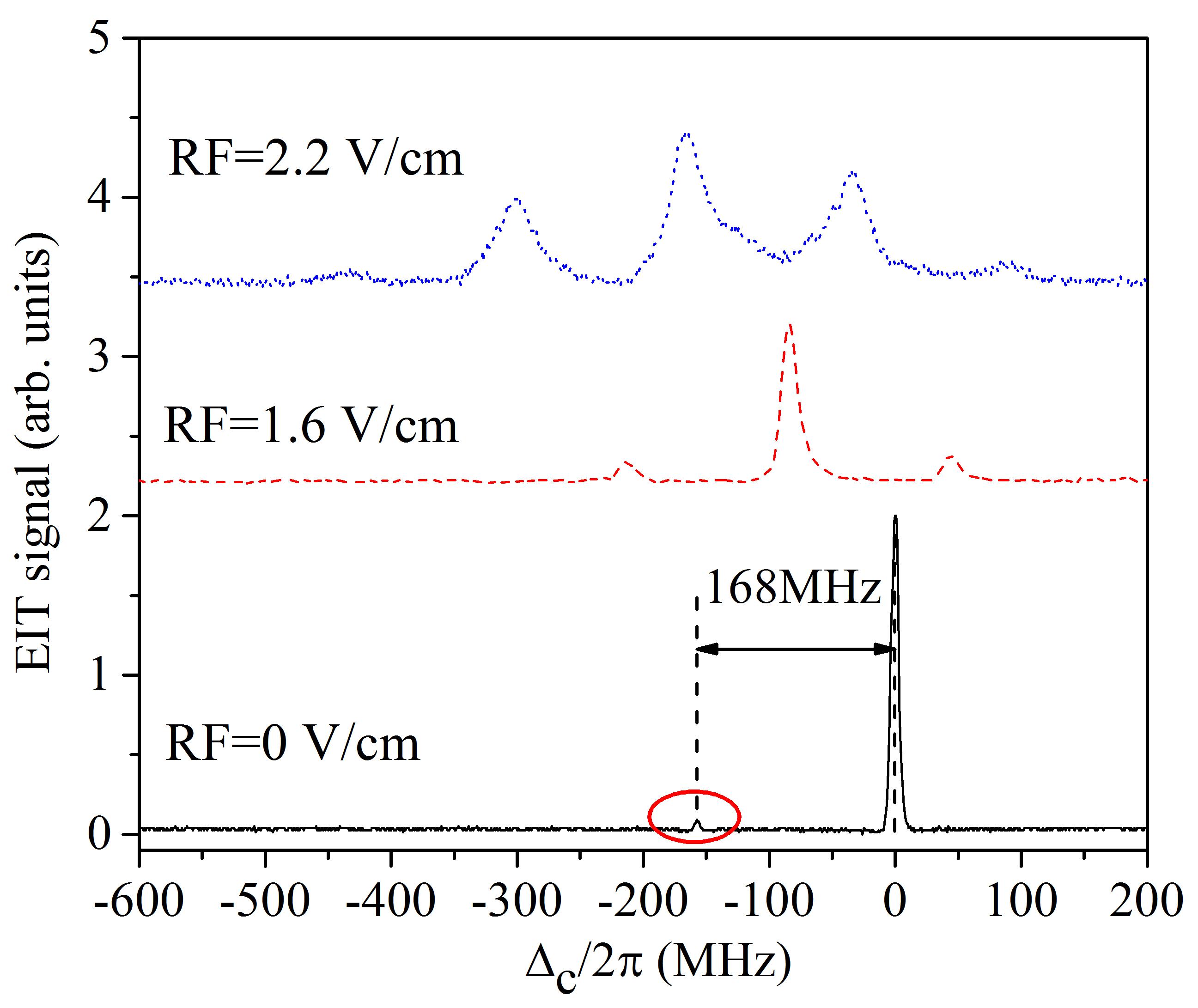}
\caption{Bottom curve: Rydberg-EIT spectrum without RF field for the 57S$_{1/2}$ Rydberg state. The dominant peak at 0 detuning corresponds to the double resonance 6S$_{1/2}(F = 4)$ $\to$ 6P$_{3/2} (F'= 5)$ $\to$ 57S$_{1/2}$, while the small peak at $-168$~MHz results from the 6S$_{1/2}(F = 4)$ $\to$ 6P$_{3/2} (F'= 4)$ $\to$ 57S$_{1/2}$ resonance.
Top curves: Rydberg-EIT spectra with harmonic RF level modulation (modulation frequency $\omega_m = 2 \pi \times$70~MHz) and the indicated RF-field amplitudes. The main EIT peak is AC Stark-shifted and develops two to four even-harmonic sidebands. The broadening of the peaks at 2.2~V/cm amplitude is due to the emergence of complex RF-induced Rydberg-level coupling with hydrogenic states.}
\end{figure}

\begin{figure*}[t]
\centering
\includegraphics[width=0.8\textwidth]{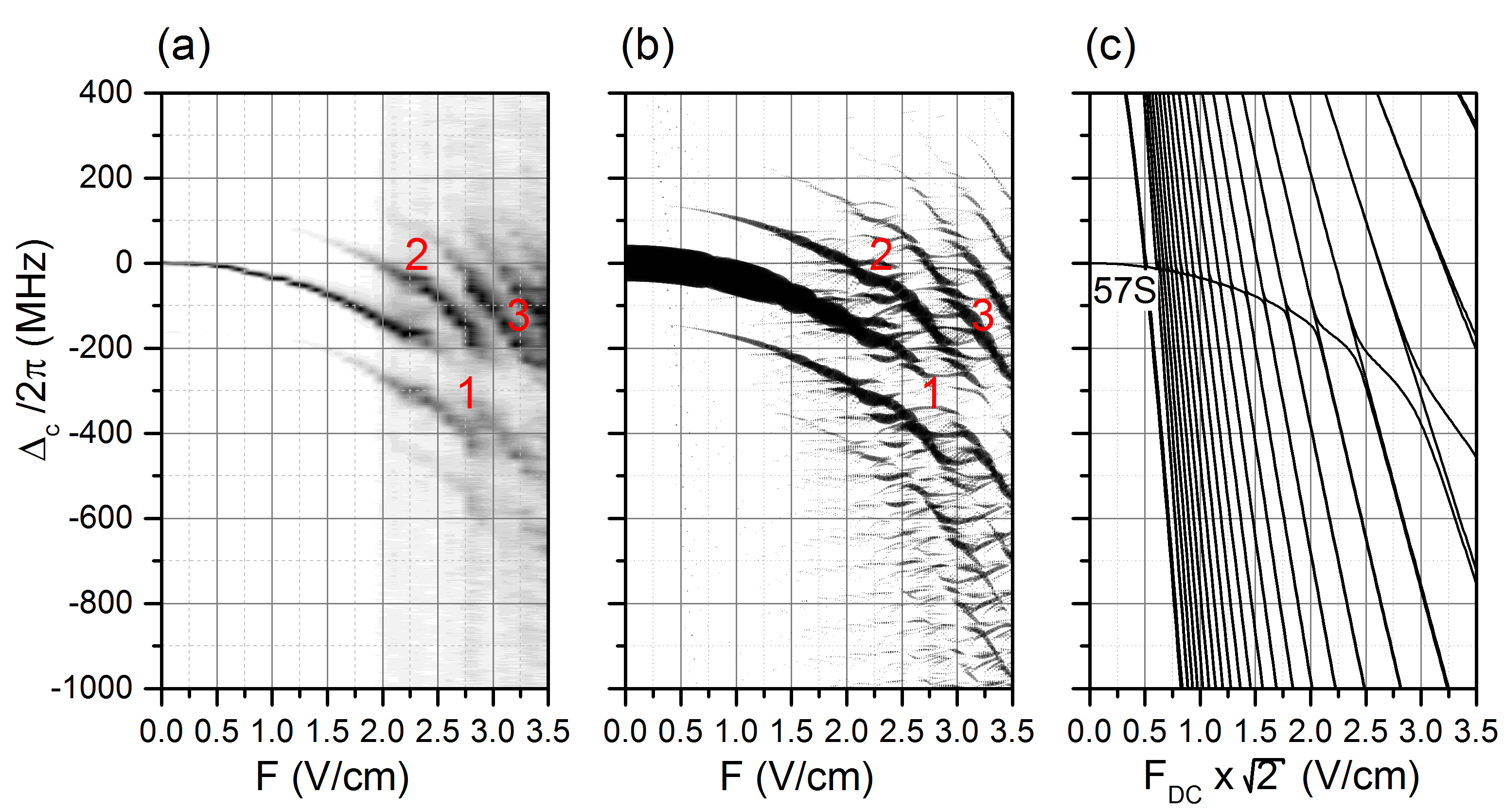}
\caption{Measurement (a) and calculation of $S_{\nu,N}$ using the Floquet model (b) of Rydberg excitation spectra as a function of RF electric-field amplitude $F$. The sinusoidal RF field modulates the Rydberg levels with a modulation frequency $\omega_m = 2 \pi \times 70~$MHz.
The labels 1-3 indicate spectral features discussed in the text. In panel (b), symbol area is proportional to line strength  $S_{\nu,N}$.
For comparison, panel (c) shows a Stark map in a DC electric field, on a field axis scaled such that the RMS fields in panels (a) - (c) match. The map shows the quadratic Stark effect of the 57$S_{1/2}$ level (polarizability $\alpha= h \times 137~$MHz/(V/cm)$^2$), a fan of hydrogenic states that exhibit linear Sark effect, and the mixing between these states.}
\end{figure*}

The top two curves in Fig.~2 show Rydberg-EIT sideband spectra with an applied sinusoidal RF signal with $\omega_{m}$ = 2$\pi \times $ 70~MHz and amplitudes $F=1.6$~V/cm (red dashed line) and $F=2.2$~V/cm (blue dotted line). It is clearly seen that in weak fields the main EIT peak (central band) is red-shifted due to the cycle-averaged AC Stark effect. Since the RF frequency is much smaller than the characteristic atomic frequency (the Kepler frequency $\approx 40$~GHz for the given state), the AC shift follows from the
DC polarizability ($\alpha= h \times 137~$MHz/(V/cm)$^2$ for 57S$_{1/2}$). For amplitudes $F=$1.6~V/cm and 2.2~V/cm the expected
AC shift of the central band, $- \alpha F^2/4$~\cite{Bason}, approximately equals $-88$ and $-166$~MHz, respectively. These estimates are in good agreement with Fig.~2. The second-harmonic sidebands, which have band indices $N = \pm 2 $, are separated by $\pm 140$~MHz from the $N=0$ band. In the limit of small $F$, the relative strengths of the $N = \pm 2 $ bands, normalized by the strength of the central band, are given by $\vert J_1(y)/J_0(y) \vert^2$, where $y=\alpha F^2/(8 \hbar \omega_m)$~\cite{Bason}. For $F=$1.6~V/cm and 2.2~V/cm one expects relative strengths of $0.11$ and $0.52$, respectively, in good qualitative agreement with Fig.~2. For the case $F=2.2$~V/cm, the line profiles of all bands, including the weak $N = \pm 4$ bands, are already significantly broadened by the RF-induced mixing of the 57S$_{1/2}$ state with the manifold of hydrogen-like states for $n=53$ (states with quantum defects $\ll 1$). This is expected from a DC Stark shift calculation shown in Fig.~3~(c),
from which it is evident that the 57S$_{1/2}$ level first intersects with hydrogen-like states at fields as low as $0.4$~V/cm and begins to substantially mix with such states between 1 and 2~V/cm.

We have performed a series of measurements such as in Fig.~2, in which we have increased the
amplitude of the sinusoidal RF modulation field from 0 to 3.7~V/cm in steps of 0.1~V/cm.
In Fig.~3~(a), the spectrum is seen to progress through several regimes that show how the oscillator strength of the $\vert e \rangle$ $\to$
$\vert r \rangle$ Rydberg transition becomes spread over the RF modulation sidebands of the 57S$_{1/2}$ level as well as other Rydberg states and their RF sidebands the 57S$_{1/2}$  level couples with.
In the weak-field regime ($F<0.7$~V/cm), the $N=0$ band is the only one clearly visible, and it exhibits the aforementioned cycle-averaged  quadratic AC Stark shift. In an intermediate regime ($0.7~$V/cm$\lesssim F \lesssim 2.5$~V/cm) the even modulation orders $N=\pm 2$ and $\pm 4$ rise in strength, while the $N=0$ order decreases in strength and drops out at about $F=2.5$~V/cm (label 1 in Fig.~3). In the weak-field model~\cite{Bason}, the dropout would occur at the field for which $J_0(y)=0$, corresponding to $y=2.40$ and a field of about $F=3.1~$V/cm in the present case. The deviation between these values for $F$ reflects the fact that at $\sim 3$~V/cm the 57S$_{1/2}$ level is already deeply within the
manifold of hydrogen-like states [see Fig.~3~(c)], and the weak-field model does not apply any more. Inspecting Figs.~3~(b) and~(c), it is seen that the RF-induced coupling with hydrogenic states pushes the $N=0$ band of the 57S-level to an energy below the corresponding 57S-line in the DC field. This effect amounts to an increase of the polarizability $\alpha$, which explains why the $N=0$ dropout occurs at a field below the value obtained from the weak-field estimate.

In the strong-field regime ($F\gtrsim 2.5$~V/cm), the shifts of the RF-induced sidebands as a function of field turn from quadratic into linear (label 3 in Fig.~3), indicating the strong RF-induced mixing of the 57S$_{1/2}$ level with RF-induced superpositions of hydrogenic states that exhibit linear Stark shifts. The undulations of the spectrum, observed below the labels 2 in Fig.~3, are a manifestation of the complex RF-induced coupling behavior of the 57S$_{1/2}$ level with the large number of such
hydrogenic states.

The absence of odd RF modulation sidebands in Fig.~3 indicates the absence of DC bias fields. In principle, DC fields
could originate from stray charges on the cell walls or from external
electric fields penetrating from outside the cell into the probe region. However,
DC fields appear to be shielded (see~\cite{Mohapatra} and Fig.~5 below).

\section{Floquet model and data interpretation}

For a quantitative
interpretation of our measured spectra, we use a non-perturbative Floquet method the accuracy of which is only limited by the size of the atomic basis set used. The method has previously been described
in Refs.~\cite{TwoPhoton, HighPower}. Here, we provide the most
relevant results.

The coupling-laser frequencies, $\omega_{\nu, N}$, at which EIT resonances are observed, and their relative excitation rates, $S_{\nu, N}$, are given by
\begin{eqnarray}
\hbar \omega_{\nu, N} & = & W_\nu + N \hbar \omega_{m} \nonumber  \\
S_{\nu, N} & = & (e F_L/ \hbar)^2 \left| \sum_k \tilde{C}_{\nu,k,N} \, {\hat{\bf{\epsilon}}} \cdot \langle k \vert \hat{\bf{r}} \vert 6P_{3/2}, m_j \rangle \right|^2, \quad
\label{eq:rates}
\end{eqnarray}
with the coupling-laser electric-field amplitude $F_L$ and polarization vector ${\hat{\bf{\epsilon}}}$.
The $\langle k \vert \hat{\bf{r}} \vert 6P_{3/2}, m_j \rangle$ are the electric-dipole matrix elements of the Rydberg basis states $\vert k \rangle = \vert n, \ell, j, m_j \rangle$ with $\vert 6P_{3/2}, m_j \rangle$.  The Floquet energies $W_\nu$ follow from the eigenphases of the time evolution operator $\hat{U}(T)$ integrated through
one period of the RF field, $T = 2 \pi / \omega_m$.
The Floquet levels $W_\nu$ have RF-dressed sidebands that are associated with the exchange of $N$ RF photons during the excitation. The Fourier coefficients $\tilde{C}_{\nu,k,N}$ are obtained from the time-dependent Floquet wave-packets, $\vert \Psi_{\nu}(t) \rangle$, and exact time-periodic functions, $C_{\nu,k}(t)$, according to
\begin{eqnarray}
\vert \Psi_{\nu}(t) \rangle & = & {\rm{e}}^{-iW_{\nu}t/\hbar} \sum_k C_{\nu,k}(t)\vert k\rangle \nonumber \\
~                           & = & {\rm{e}}^{-iW_{\nu}t/\hbar} \sum_{k}\sum^{\infty}_{N=-\infty} \tilde{C}_{\nu,k,N}{\rm{e}}^{-i N \omega_{m} t}\vert k\rangle, \nonumber \\
\tilde{C}_{\nu,k,N}&=&\frac{1}{T}\int_{0}^{T} C_{\nu,k}(t){\rm{e}}^{iN\omega_{m}t}dt,  \quad
\label{eq:rates2}
\end{eqnarray}
For more details, see~\cite{HighPower}.

The basis $\{ \vert k \rangle \}$ must be chosen large enough to cover important couplings. Since the theory has to accurately describe mixing with hydrogenic states, it is obvious that the basis has to  include all $\ell$ and $j$-values, which are $\ell=0, ..., n-1$ and $j=\ell \pm 1/2$, respectively (for $\ell=0$ only $j=1/2$). Since the fields are $\pi$-polarized, $m_j=1/2$. We have found that the basis can be limited to states with effective principal quantum numbers $51.1\le n_{\rm{eff}} \le 54.9$. In the method, the time evolution operator $\hat{U}(T)$ is integrated in time steps $\Delta t$ through one RF cycle. The value of $\Delta t$ must be small enough that $\vert W_k - W_{k'} \vert \Delta t / \hbar \le 2 \pi$  for all energy differences $W_k - W_{k'}$ for basis states  $\vert k \rangle $ and $ \vert k' \rangle$. In the present case, $\Delta t = T/2048$ to $T/4192$ is appropriate.

The agreement between experimental and calculated excitation rates, seen in Fig.~3, is generally very good, both in terms of the positions of spectral features as well as relative signal strengths. The beading within the steep bands in the experimental data reflects the step size in the electric-field change between the scans. Comparing Figs.~3~(a) and~(b) we note agreement even in fine details, such as the significant kink of the $N=2$ band at 2.3~V/cm (just below label 2) and the groups of levels that form  bridges between the $N=0$ and $N=-2$ bands at 3.0~V/cm (to the lower-right of label 1). In the high-field regime, most of the spectral features are not due to single, isolated levels (Floquet states and RF sidebands), but due to signals associated with several levels merging into compound structures. Comparing calculated with experimental spectroscopic Floquet maps, we estimate the deviation between the experimental and theoretical electric-field axes to be below $\sim 0.1~$V/cm.
The uncertainty may be due to field inhomogeneities in the Cs vapor cell or a calibration uncertainty of the RF-voltage measurement of the signal applied to the field plate (see Fig.~1).

\begin{figure} [t]
\vspace{-1ex}
\centering
\includegraphics[width=0.5\textwidth]{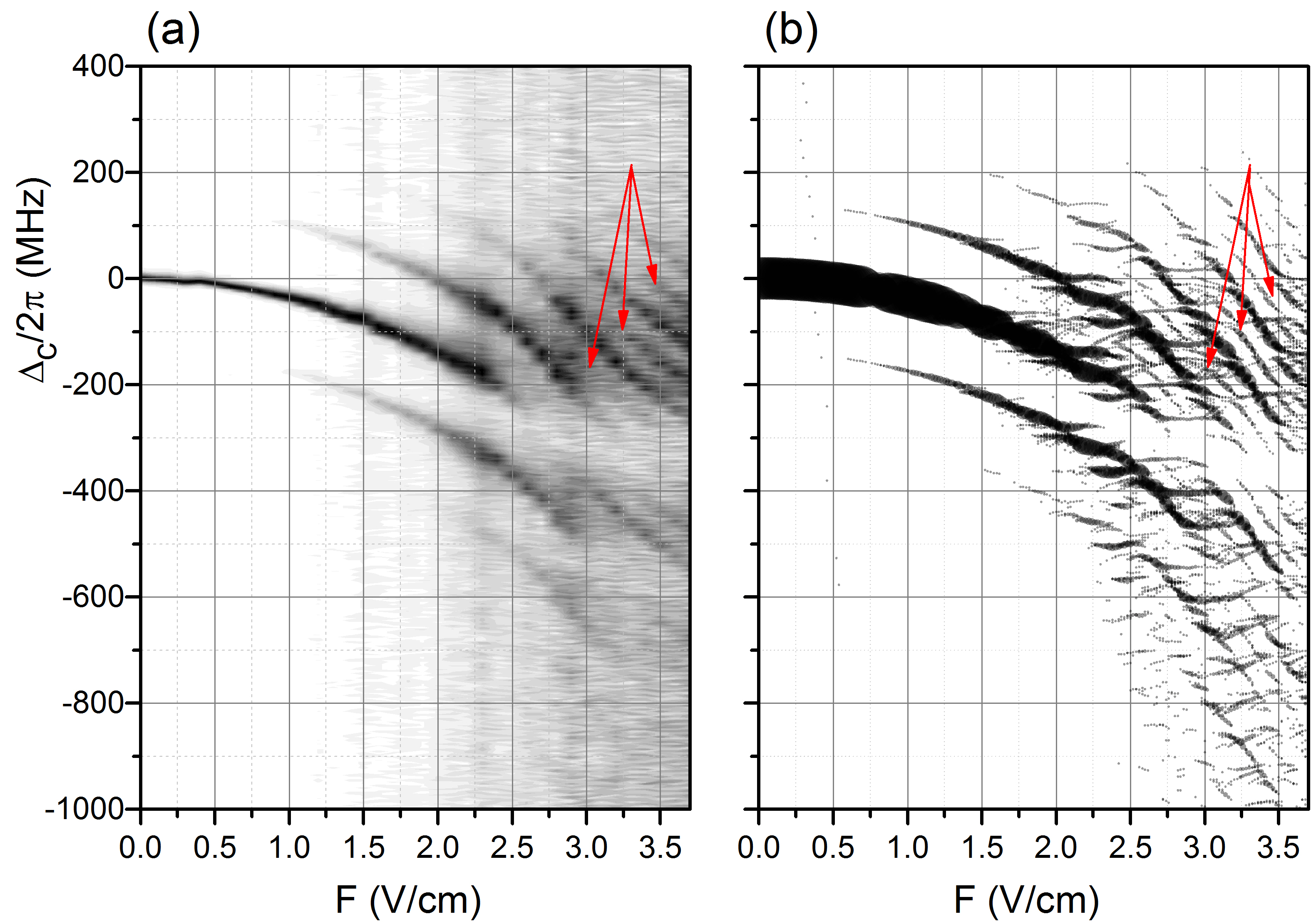}
\vspace{-1ex}
\caption{Measurements and calculations analogous with Fig.~3, but using a sightly anharmonic waveform.
To enhance the signal at high RF field, the probe Rabi frequency for panel (a)
has been chosen 1.5 times as large as in Fig.~3.
The calculation in (b) is for an RF field $F_{RF}(t) = F [\cos(\omega_m t)+ 0.025 \cos(2 \omega_m t)]$
with $\omega_m = 2 \pi \times 70$~MHz.}
\end{figure}

We also have investigated modulation with a distorted RF signal that has a small degree of anharmonicity.
The result of this measurement, also performed at a modulation frequency of 70~MHz,
is shown in Fig.~4(a).
In comparison with Fig.~3, where the applied RF modulation was a pure sine wave, the spectra in Fig.~4 exhibit additional anharmonicity-induced odd-order modulation bands, labeled with arrows.
The odd sidebands apparently result from higher-harmonic components in the distorted harmonic RF field.
To verify this assumption, we have computed numerous Floquet spectra for periodic RF fields with expansion
$F(t) = F \, [ \sum_{n=0}^{3} \alpha_n \cos (n \omega_m t) + \sum_{m=1}^{3} \beta_m \sin( m \omega_m t)]$. We set $\alpha_1 = 1$ to
represent the main RF term and keep all other Fourier coefficients small. In view of earlier work~\cite{Mohapatra} and Fig.~5 below, we expect that higher harmonics rather than the DC component of the distorted RF signal will give rise to the additional sidebands. We have therefore focused on cases with $\alpha_0 = 0$ (which have a zero cycle-averaged field).
In Figure 4~(b) we show a calculated Floquet spectrum for $\alpha_2 = 0.025$. Both in experiment and calculation, the anharmonicity-induced odd-order modulation sidebands begin to emerge at about 2.5~V/cm and then quickly rise in visibility.
At the high-field limit in Fig.~4, odd and even sidebands have similar visibility.
Figure~4 and similar calculations demonstrate that RF modulation spectroscopy of Rydberg atoms is quite sensitive to anharmonic content in periodic RF waveforms, and that the susceptibility of the atomic response to the anharmonicity increases with the RF field.

\begin{figure}[t]
\vspace{-1ex}
\centering
\vspace{-1ex}
\includegraphics[width=0.5\textwidth]{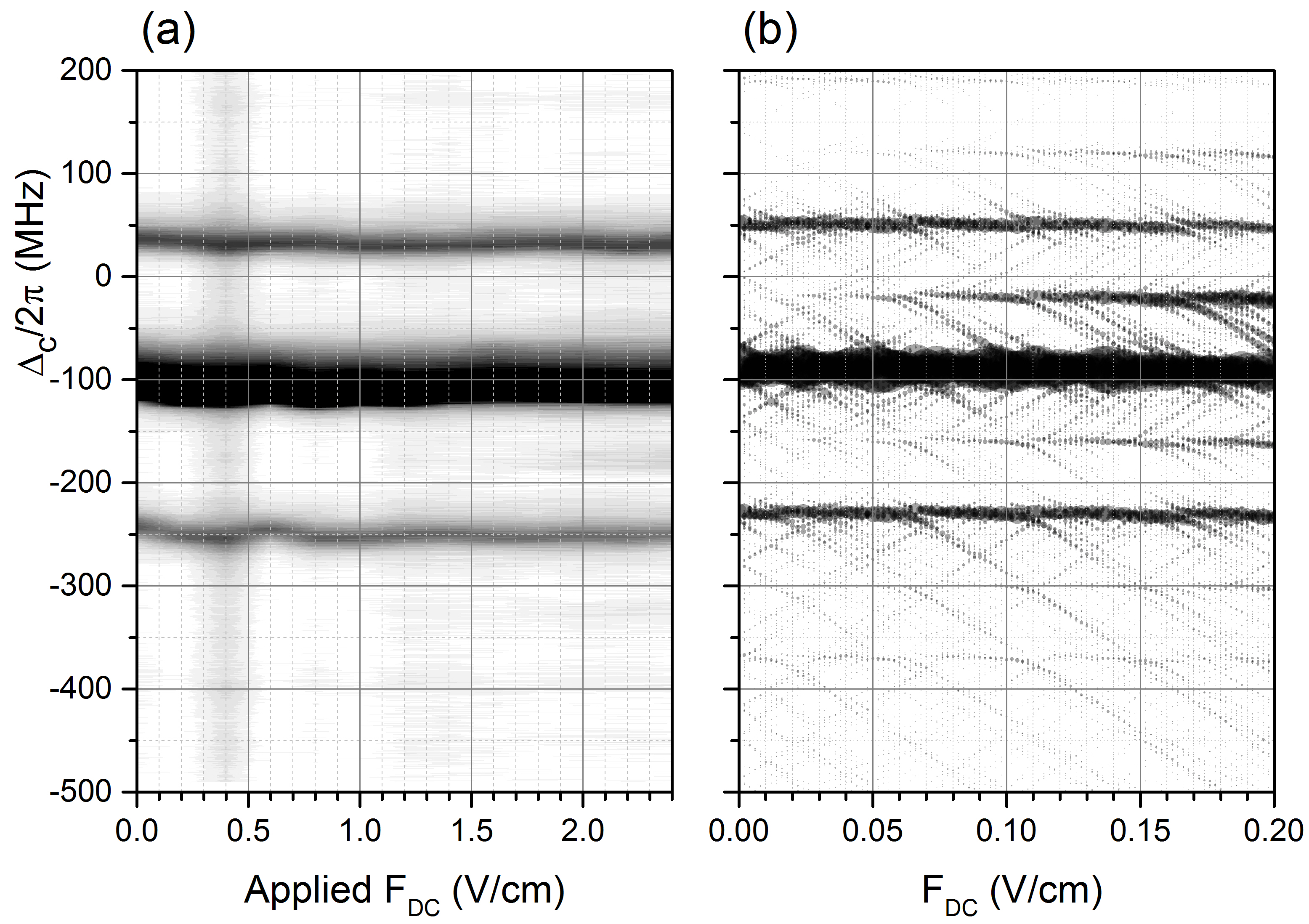}
\caption{Experimental (a) and calculated (b) Rydberg-EIT spectra as a function of a DC offset field, $F_{DC}$, added to
a purely harmonic RF modulation with frequency $\omega_m = 2 \pi \times 70$~MHz and fixed amplitude $F=1.6$~V/cm. In panel (a) the $x$-axis shows the applied DC field, which is much larger than the shielded, remaining DC field within the atom-field interaction region.
To enhance sensitivity for weak odd-band signals, the probe Rabi frequency has been chosen 1.5 times as large as in Fig.~3.}
\end{figure}

Finally, we have analyzed the effect of a DC offset field on Floquet maps of the RF-modulated 57$S_{1/2}$ Rydberg state.  The directions of the applied DC and RF electric fields are parallel to each other, as evident from Figure~1. In the calculation shown in Fig.~5~(b),
over the range of $F_{DC}$ from 0 to 0.2~V/cm we see the strength of the first-order ($N= \pm 1$) sidebands rise to about 10~$\%$ of the strength of the $N=0$ band, in qualitative agreement with the low-field model in~\cite{Bason}. The DC Stark shift, $-\alpha F_{DC}^2/2$, is very minor and near-invisible in Figure~5~(b). However, we do not see the first-order sidebands in the experimental Floquet map shown in Figure 5~(a). The fact that an applied DC electric field is not effective in generating any visible atomic response can be attributed to a DC shielding effect, in which the DC electric field is screened by surface charges on the inside cell walls. The shielding effect, which has been reported before Ref.~\cite{Mohapatra}, is attributed to ions and electrons produced by ionization of Rydberg atoms and photo-electric effect on the cell walls. Under presence of an additional RF field (our case), the shielding effect persists, showing that the high-frequency field does not impede the DC shielding. From Fig.~5 we find a lower bound of the DC shielding factor of $\sim 100$, limited by experimental sensitivity and the low value of the applied DC field. The actual shielding factor may be much higher.

\section{Conclusion}

In summary, we have studied RF-modulated Rydberg-EIT spectra in a cascade three-level system in a cesium room-temperature vapor cell. The observed Rydberg Floquet spectra exhibit field-induced level shifts and sidebands in weak RF fields, whereas higher-order sidebands, complex state mixing and level crossings with hydrogenic states are found in strong RF fields.
Weak anharmonicity of the RF signal has been found to produce odd modulation sidebands. In work not shown we have obtained qualitatively similar results for modulation frequencies in the range from 40 to 80~MHz.
The experimental results are well explained by our Floquet model, even considering
details of the complex high-field spectra. The Rydberg-EIT spectroscopy presented here could be applied in an atom-based technique for antenna-free measurement of RF electric fields. The method allows for fast, all-optical readout of the RF-modified Rydberg level structure. In order to evaluate whether the Rydberg-EIT method allows spectral analysis of periodic RF waveforms, a detailed study using a wide range of synthesized periodic signals may be conducted.

The work was supported by the 973 Program (Grant No. 2012CB921603), NNSF of China (Grants No. 11274209, 61475090, 61378039, and 61378013), and Research Project Supported by Shanxi Scholarship Council of China (2014-009). GR acknowledges support by the NSF (PHY-1506093) and BAIREN plan of Shanxi province.

\end{document}